\documentclass{ptapap}
\usepackage{amsmath}
\usepackage{amssymb}

\author{Igor Soszy\'nski}[OAUW]
\affil[OAUW]{Astronomical Observatory, University of Warsaw, Al.~Ujazdowskie~4, 00-478~Warszawa, Poland}

\title{Eppur si muove...\\On the Origin of Long Secondary Periods in Red Giant Stars}

\headtitle{Long Secondary Periods in Red Giant Stars}

\begin{document}

\maketitle

\begin{abstract}

Long secondary periods (LSPs), observed in a third of pulsating red giant and supergiant stars, are the only unexplained type of large-amplitude stellar variability known at this time. Numerous authors have explored various scenarios for the origin of LSPs, but were unable to give a final solution to this problem. We present known properties of LSP variables and show new results proving that the physical mechanism responsible for LSPs is binarity. Namely, the LSP light changes are due to the presence of a dusty cloud orbiting the red giant together with a brown-dwarf companion and obscuring the star once per orbit. In this scenario, the low-mass companion is a former planet that accreted a significant amount of mass from the envelope of its host star and grew into a brown dwarf.

\end{abstract}

\section{Introduction}

Pulsating red giant\footnote{In this paper we use the term ``red giant'' for both first-ascent red giant branch (RGB) and asymptotic giant branch (AGB) stars.} and supergiant stars are collectively called long-period variables (LPVs). This is a very numerous but still poorly understood class of variable stars. Hundreds of thousands of LPVs have been discovered in the Optical Gravitational Lensing Experiment (OGLE) photometric databases \citep{soszynski2009,soszynski2011,soszynski2013}. The most spectacular and the best known subtype of LPVs are Mira variables. These are single-mode pulsating AGB stars with periods ranging from about 80 to over 1000 days and amplitudes reaching 9~magnitudes at visual wavelengths. Stars belonging to the second subclass of LPVs -- semiregular variables (SRVs) -- are typically double-mode pulsators with periods from about 30 to 500 days. As the name suggests, their light curves exhibit both periodic and irregular behavior. Finally, the most numerous group of LPV are OGLE small-amplitude red giants\footnote{The term ``OGLE small-amplitude red giant'' was coined by Laurent Eyer and used for the first time in the paper by \citet{wray2004}.} (OSARGs) -- multiperiodic cool giants, usually oscillating simultaneously in three, four, or even more radial and non-radial modes.

Approximately one third of LPVs exhibit additional periodic modulation called long secondary periods (LSPs). This phenomenon has been observed in red giant and supergiant stars for many decades \citep[e.g.][]{oconnell1933,payne1954,houk1963}, but the origin of the LSPs remained a mystery. Many possible explanations have been considered by various authors: radial or non-radial stellar pulsation, turnover of giant convective cells, episodic dust ejection, magnetic activity, or binarity. In recent years, two of these mechanisms have been favored in the literature: the non-radial oscillatory convective modes confined to the outermost layers of the red giant star \citep[e.g.][]{saio2015,takayama2020} and the presence of a dusty cloud orbiting the red giant with a low-mass companion in a close orbit \citep[e.g.][]{soszynski2014,soszynski2021}.

In this paper, we present some basic properties of LSP variables and discuss arguments in favor and against the binary explanation of the LSP phenomenon. In particular, we report our latest findings that strongly support the binary scenario. We studied mid-infrared time-series photometry extracted from the Near-Earth Object WISE Reactivation Mission (NEOWISE) archive and found that a large fraction of the infrared LSP light curves exhibit both primary and secondary minima (that can be interpreted as eclipses), while the visual light curves show only one minimum per the LSP cycle. Details of our investigation can be found in the paper by \citet{soszynski2021}.

\begin{figure}
\includegraphics[width=\textwidth]{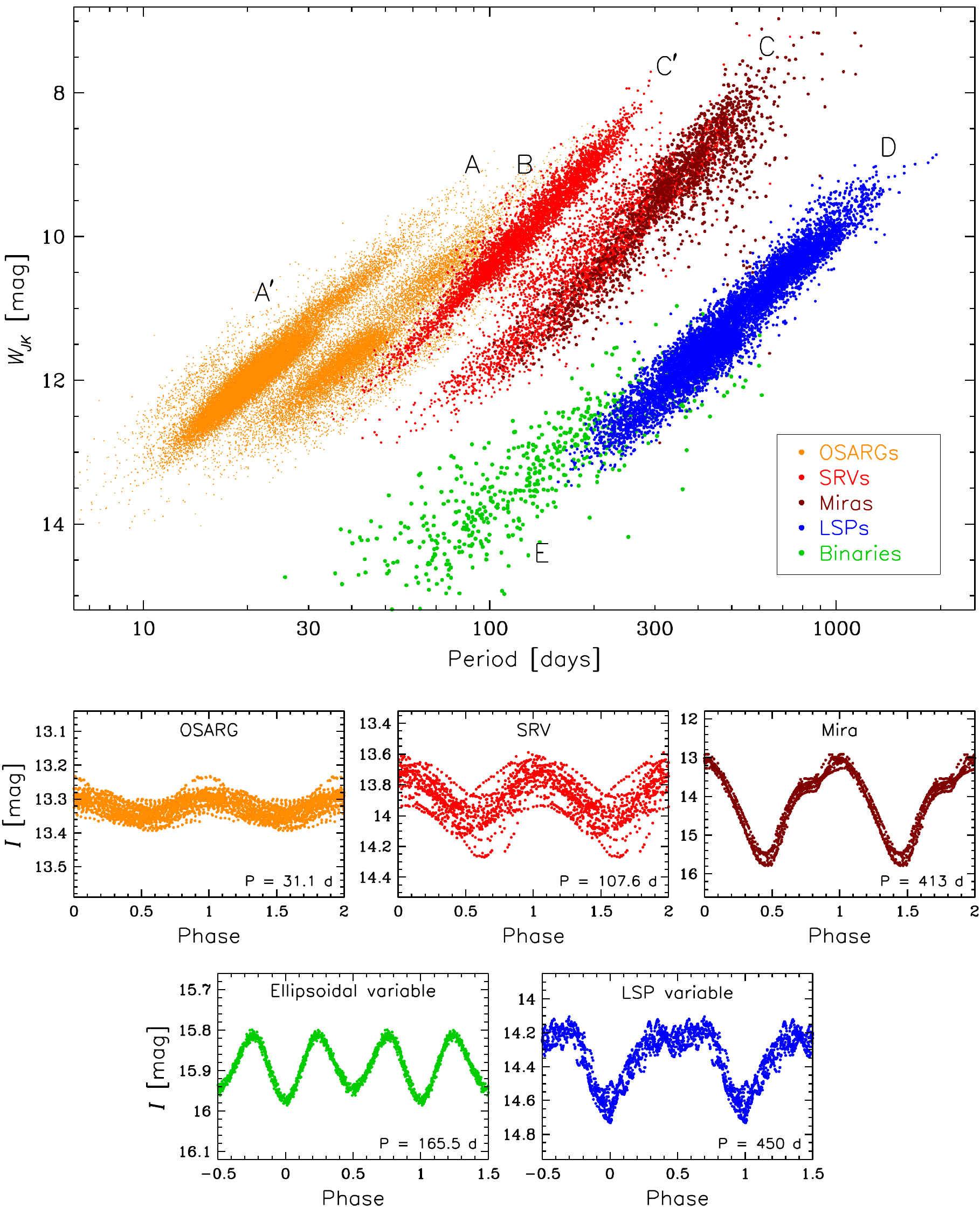}
\caption{Upper panel: period--Wesenheit index diagram for LPVs in the Large Magellanic Cloud. The near-infrared Wesenheit index is an extinction-free quantity defined as $W_{JK}=K_\mathrm{s}-0.686(J-K_\mathrm{s})$, where {\it J}- and $K_\mathrm{s}$-band magnitudes originate from the IRSF Catalog \citep{kato2007}. Different colors of the points indicate different types of LPVs: orange -- OSARGs, red -- SRVs, brown -- Miras, blue -- LSP variables, and green -- eclipsing and ellipsoidal close binary systems containing a red giant as one of the components. Lower panels: typical {\it I}-band light curves of different types of LPVs.}
\label{fig1}
\end{figure}

\section{Properties of LSP Variables}

LPVs follow a series of sequences \citep{wood1999} in the infrared period--luminosity (PL) or period--Wesenheit diagram (upper panel of Fig.~\ref{fig1}). Mira variables lie on sequence C, associated with the fundamental mode of stellar pulsation, SRVs populate sequences~C and~C$'$, corresponding to the fundamental and first-overtone modes, while OSARGs are preferentially located in sequences~A, A$'$, and B. Close eclipsing or ellipsoidal binary systems containing a red giant as one of the components form broad but clearly defined sequence~E in the lower part of the PL diagram. Finally, sequence~D hosts giants exhibiting LSPs and therefore these objects are sometimes referred to as ``sequence~D stars''.

The LSP modulation is detectable in a large fraction (30\% or more) of SRVs and OSARGs. It is unclear whether this phenomenon occurs also in Mira variables. LSPs range from several months to several years, roughly 10 times longer than the primary pulsation periods which usually fall into PL sequences B, C$'$, or in the gap between these two sequences. \citet{mcdonald2019} linked this region of the red giant PL plane with the onset of substantial mass-loss via dust-driven winds. Indeed, \citet{wood2009} found that LSP stars have a mid-infrared flux excess due to absorption of stellar light by circumstellar dust followed by re-radiation in the infrared range. Moreover, this dust appears to be in a non-spherical distribution around the giant. These conclusions were supported by \citet{pawlak2021}, who analyzed the spectral energy distribution of a large number of LSP and non-LSP giants.

The light curves associated with the LSPs have a distinctly different morphology than the light curves produced by stellar pulsation in red giants (compare the light curves of a typical Mira, SRV, and OSARG with the light curve of an LSP variable in the lower panels of Fig.~\ref{fig1}). In the well-defined sequence~D stars, the visual light curve can be divided into two parts, each lasting approximately half of the LSP cycle \citep{soszynski2014}. During the first part, the brightness does not change at all (except short-period stellar oscillations) or it increases slowly with time. The other half of the LSP light curve has a triangular shape: the brightness decreases, reaches a minimum, and then increases. Such light curves resemble those of eclipsing binary systems with one very broad eclipse per orbital period. We emphasize that this description refers to the LSP light curves in the visual passbands, including the {\it I}-band used by the OGLE project. In Section~3, we will demonstrate that the infrared light curves differ in one detail: many of them exhibit additional minima that can be interpreted as the secondary eclipses.

Radial velocities have been measured for several dozen LSP variables \citep{hinkle2002,wood2004,nicholls2009}. The peak-to-peak amplitudes of the velocity curves range from about 2~to 7\;km\;s$^{-1}$, with a median value of 3.5\;km\;s$^{-1}$. In the binary model, such velocity amplitudes indicate a very-low-mass stellar or a brown-dwarf companion (with a mass in the range $\sim$0.06--0.12\;$\textrm{M}_\odot$) in a close orbit around the giant star. The radial velocity curves of LSP variables are usually non-sinusoidal and moreover have similar shapes in different objects. We will return to this point in the next section, as this is one of the arguments that led \citet{wood2004} and other authors to reject binarity as a valid explanation of the LSP phenomenon.

The spectroscopic observations were also used to measure changes in the effective temperature during the LSP cycle \citep{wood2004}. The amplitudes of the temperature changes turned out to be very small (if any) -- much smaller than predicted by a model linking the LSPs with the radial pulsation modes excited in the giant stars.

\section{The Origin of LSPs in Red Giant Stars}

\citet{wood1999} were the first who suggested that the LSPs may result from obscurations of the red giant star by a cloud of dust associated with an orbiting companion. Strong support for this hypothesis was provided by \citet{soszynski2004}, who noticed that sequence~D in the PL diagram partially overlaps with sequence~E formed by close binary systems and can be interpreted as its extension (Fig.~\ref{fig1}). However, \citet{wood2004} ruled out the binary explanation and opted for the non-radial pulsation modes as a possible cause for the LSPs. \citet{wood2004} and later \citet{nicholls2009} argued that the observed asymmetric radial velocity curves would indicate eccentric orbits of the binary systems. However, similar shapes of the velocity curves observed in different LSP stars imply similar angles of periastron, which is unlikely, because one may expect that randomly aligned orbits should have angles of periastron uniformly distributed between 0 and 2$\pi$.

An additional argument against the binary scenario was the large fraction of LSP variables among red giants, which is in conflict with the so-called ``brown dwarf desert'' -- an observed deficit of brown dwarf companions to main sequence stars (that are progenitors of red giants). \citet{nicholls2009} estimated that less than 1\% of main sequence stars have a low-mass (0.06--0.12\;$\textrm{M}_\odot$) companion, which is in sharp contrast to the large incidence rate (at least 30\%) of LSP stars.

\begin{figure}
\includegraphics[width=\textwidth]{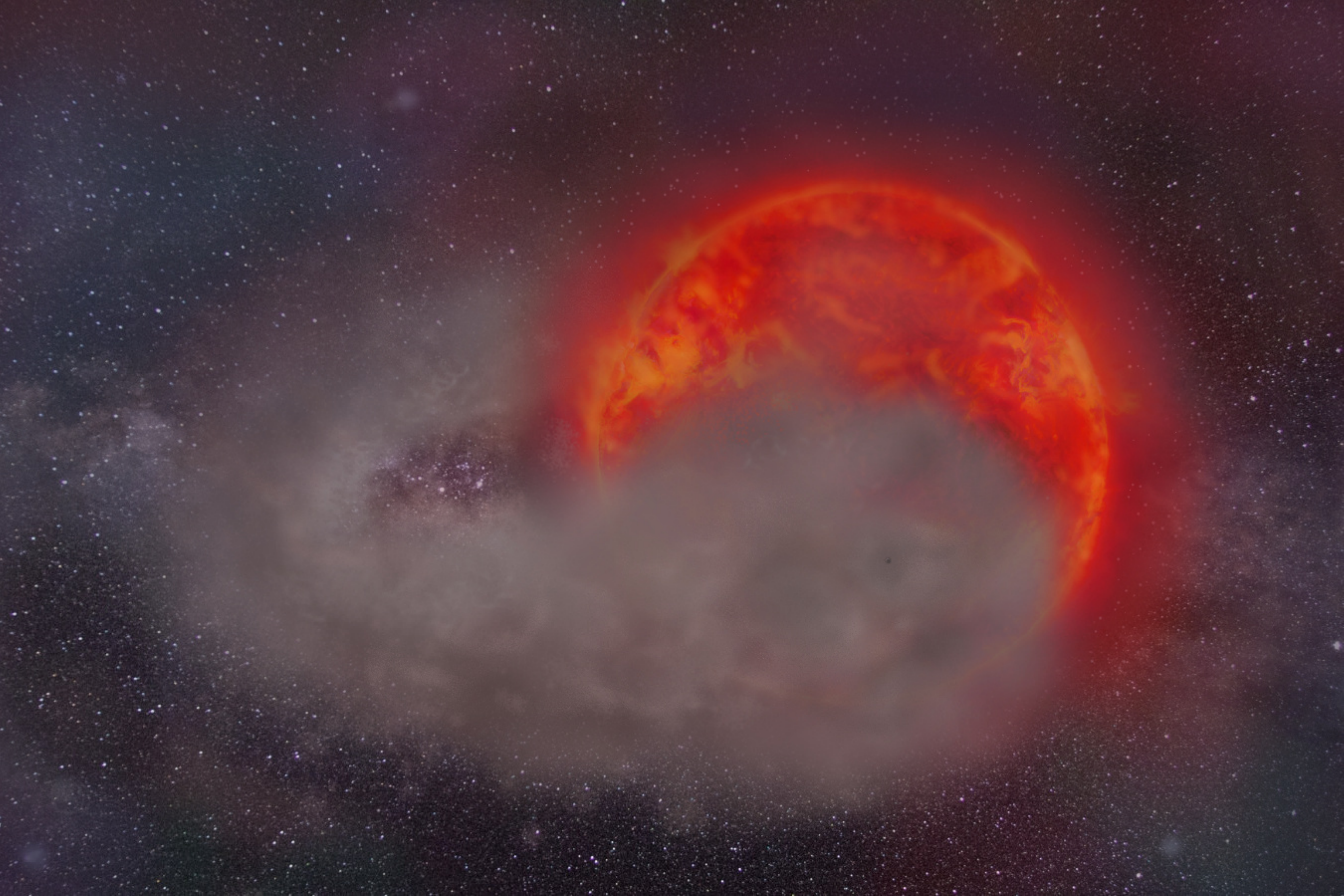}
\caption{Artistic impression of a red giant star obscured by a comet-like dusty cloud surrounding a low-mass companion. Author: Matylda Soszyńska.}
\label{fig2}
\end{figure}

These arguments against the binary hypothesis were addressed by \citet{soszynski2014}. In the binary model of the LSP phenomenon, the low-mass companion is submerged in an extended cloud of dust orbiting the red giant just above its surface. An artistic impression of such a system is shown in Fig.~\ref{fig2}. Three-dimensional hydrodynamical simulations of interacting binary systems containing an AGB star \citep[e.g.][]{theuns1993,saladino2019,bermudez2020} predict that the material lost by the giant via stellar wind should form such a spiral cloud that obscures the red giant once per orbit causing the triangular dips in the LSP light curves.

\citet{soszynski2014} considered a simple model of a dusty cloud with a comet-like tail orbiting the red giant star with a low-mass companion in a close circular orbit. It turned out that such a model can successfully reproduce both the characteristic shapes of the LSP light curves and the non-sinusoidal radial velocity curves. The asymmetric velocity curves can be produced in binary systems with circular orbits due to the Rossiter-McLaughlin effect. In brief, a dusty cloud associated with the low-mass companion may block part of the light emitted by the approaching or receding parts of the red giant disk in different orbital phases, which results in a non-sinusoidal velocity curve. This mechanism explains why the LSP velocity curves are asymmetric and similar to each other.

Recently, \citet{soszynski2021} reported a discovery that -- as it seems -- finally solved the mystery of the origin of LSPs in red giant stars. The obvious consequence of the binary hypothesis is that the energy absorbed by the dusty cloud over a wide range of the electromagnetic spectrum should be reemitted in the infrared wavebands. Thus, one can expect that the infrared light curves of LSP variables should exhibit secondary eclipses during the cloud's passage behind the red giant.

Our procedure was as follows. In the first stage, we performed a search for LSP variables in the photometric database obtained by the OGLE survey \citep{udalski2015}. Currently, OGLE utilizes the 1.3-m Warsaw telescope at Las Campanas Observatory, Chile, to monitor the brightness of about 2~billion stars in the Galactic bulge, Galactic disk, and Magellanic Clouds. The long-term {\it I}- and {\it V}-band time-series photometry is ideally suited for detecting and studying variable red giant stars. One of the results of the second and third phases of the OGLE survey (carried out in 1997--2009) were catalogs of about 340\;000 LPVs in the Milky Way and Magellanic Clouds \citep{soszynski2009,soszynski2011,soszynski2013}. We extracted the OGLE-IV photometry (gathered between 2010 and 2020) of these objects and isolated well-defined LSP variables from this sample. Our classification was primarily based on the characteristic light curve shapes of the LSP stars and their position on sequence~D in the PL diagram (Fig.~\ref{fig1}). Additionally, we performed a limited search for LSP variables in the region of the sky monitored by the OGLE-IV project but not observed during the previous phases of the survey. As a result, we obtained a sample of about 16\;000 sequence~D stars ($\sim$9000 in the Milky Way and $\sim$7000 in the Magellanic Clouds) with relatively large amplitudes of the LSP modulation.

In the second stage of our procedure, we crossmatched our sample of LSP stars to the NEOWISE archive \citep{mainzer2014}. These mid-infrared time-series data have been gathered by the Wide-field Infrared Survey Explorer \citep[WISE;][]{wright2010} -- a 40-cm space telescope launched in 2009. Since 2013, the NEOWISE mission surveys the entire sky in two mid-infrared filters centered at 3.4~$\mu$m (labeled W1) and 4.6~$\mu$m (W2). Each sky location is visited by the WISE telescope every six months, thus typically around 12 epochs per star have been collected until the NEOWISE 2020 Data Release. The exception to this rule are regions near the ecliptic poles, which were observed much more frequently due to a Sun-synchronous polar orbit of the WISE satellite. Fortunately, an area around the South Ecliptic Pole was regularly observed by the OGLE project, because it is located in the outskirts of the Large Magellanic Cloud, so a number of our LSP variables had well-sampled infrared light curves from the NEOWISE project.

\begin{figure}[t]
\includegraphics[width=\textwidth]{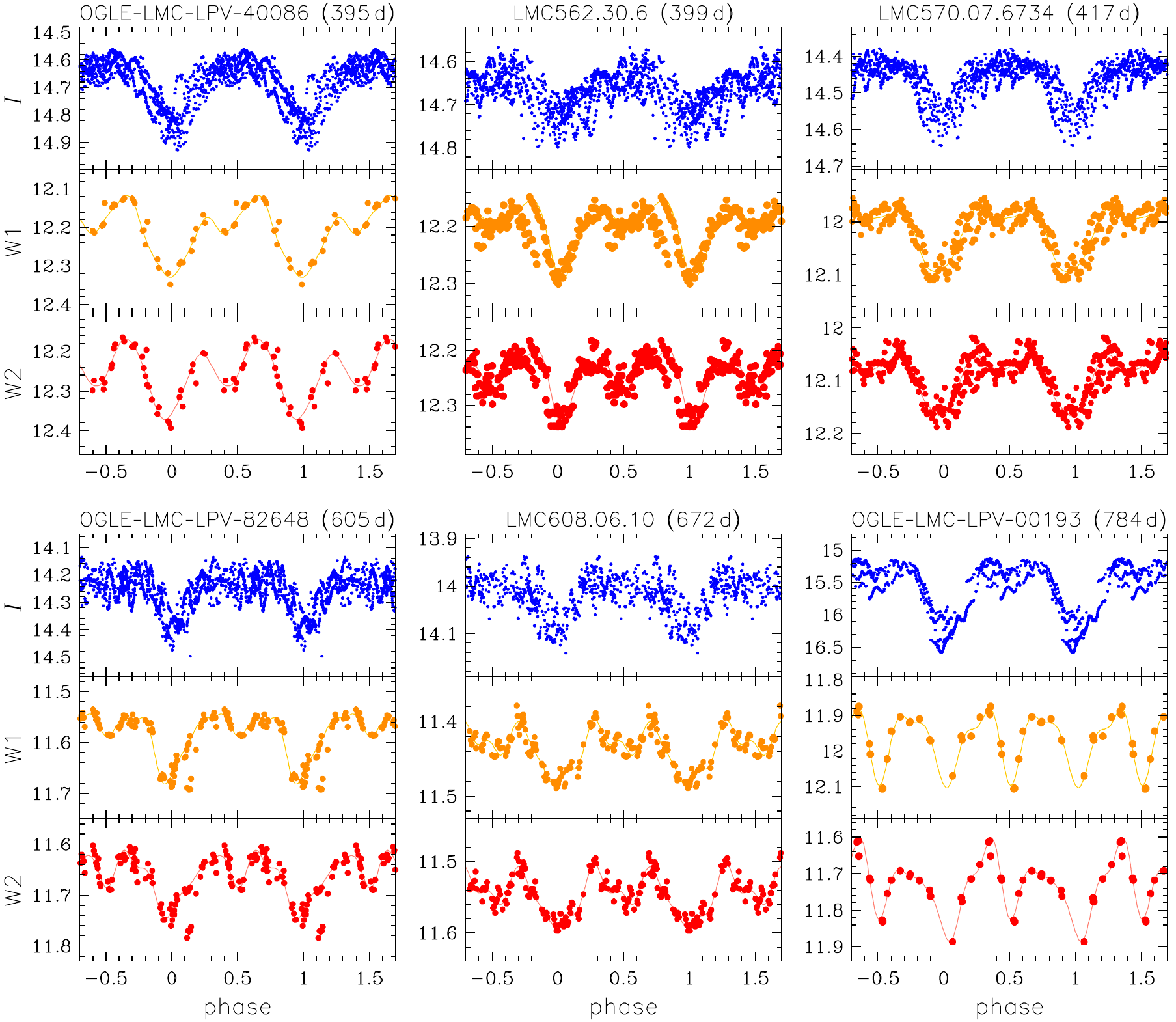}
\caption{Folded light curves of six LSP variables from the Large Magellanic Cloud. In each panel, blue, orange, and red points show OGLE {\it I}-band, NEOWISE W1, and NEOWISE W2 photometry, respectively. The primary minima visible in all wavebands (at phases 0 and 1) are caused by the eclipses of the red giant by an orbiting cloud of dust. The secondary minima, visible only in the infrared light curves (at phase 0.5), occur when the cloud is hidden behind the red giant.}
\label{fig3}
\end{figure}

In the last stage of our analysis, we visually inspected the NEOWISE light curves of our sequence~D stars. In most cases, the phase coverage was too sparse to draw definite conclusions about the morphology of the infrared time series. Additionally, we rejected objects with low or variable amplitudes of the LSP modulation or stars with exceptionally large amplitudes of the stellar pulsations that masked the LSP modulation. Our final set of LSP variables with well-sampled, large-amplitude, and stable infrared light curves contained about 700 objects. Fig.~\ref{fig3} shows the OGLE {\it I}-band and NEOWISE W1- and W2-band light curves of six LSP variables from this sample.

As one can see, the optical and infrared LSP light curves have similar characteristics. The triangular-shaped primary minima lasting about a half of the LSP cycle are visible in both optical and infrared passbands. The most striking difference between OGLE and NEOWISE light curves can be seen at phase 0.5. Approximately half of the LSP stars show pronounced secondary dips visible only in the infrared range.

\section{Discussion and Conclusions}

The secondary minima visible in the infrared wavelengths confirm that the LSP variability is caused by eclipses from a comet-like cloud of dust surrounding a low-mass companion. As this cloud passes between us and the red giant star, we observe the broad primary eclipse in the optical and infrared light. When the cloud and companion then pass behind the giant, their infrared emission vanishes, causing the secondary eclipse observed at infrared wavelengths only. Thus, we have likely found a solution to the decades-old mystery of the LSP variability in red giant stars.

The widths and depths of the secondary eclipses in relation to the primary ones vary significantly from star to star (Fig.~\ref{fig3}). Note for example that in some cases the secondary dips are much narrower than the primary ones. This indicates that the densest and hottest part of the dusty cloud (that emits the bulk of the infrared light) is significantly smaller than the whole cloud with its comet-like tail. This explains why the secondary eclipses are not visible in all well-sampled infrared light curves of LSP stars. Generally, the shapes of the secondary eclipses in various photometric passbands depend on many factors: the morphology of the orbiting cloud of dust, the temperature and density distributions inside this cloud, its transparency in various photometric bands, and the inclination of its orbit. As a result, the secondary eclipses may be too shallow to be detectable in some LSP stars, which we actually noticed. 

Careful examination of the secondary eclipses in various filters can unveil the properties of LSP variables. For example, in most cases the secondary minima are located symmetrically between the primary minima, which indicates circular orbits of the low-mass companions. Also, the relative depths of the secondary eclipses in the W1 and W2 bands can be used to estimate the temperature of the dusty cloud orbiting the giant star.

The last problem that needs to be addressed is the high prevalence ($\sim$30\%) of sequence~D stars among LPVs, which implies a large fraction of red giants with a low-mass companion on a circular orbit. To solve this problem it should be assumed that the companion is a former planet that accreted a significant amount of mass from the envelope of its host star and appears as a brown dwarf or even a low-mass star. It is known that planets detected around red giants are on average much more massive than those around main-sequence and subgiant stars \citep[e.g.][]{jones2014,niedzielski2015}, suggesting that planets indeed grow via accretion from the stellar wind of their hosts. Also, the nonspherical shapes of the majority of planetary nebulae are usually explained by the presence of stellar or substellar objects that survived the AGB phase of their companions \citep[e.g.][]{decin2020}. Thus, there are strong observational arguments supporting the hypothesis that LSPs in red giant stars originate from former planets that accumulated mass from expanding envelopes of their hosts.

\begin{figure}
\includegraphics[width=\textwidth]{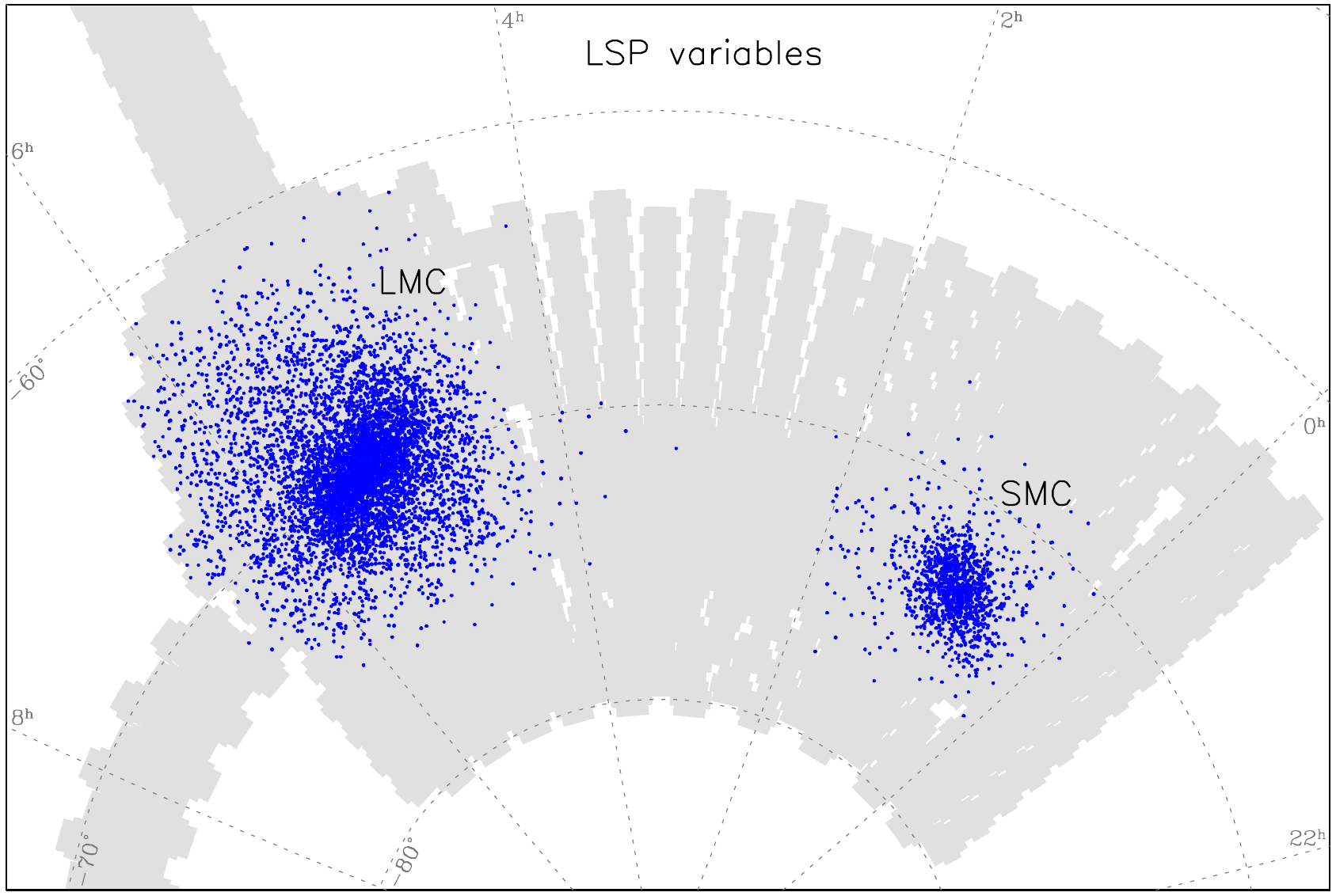}
\caption{On-sky distribution of about 7000 LSP variables in the Magellanic Clouds. The gray area shows the OGLE-IV footprint.}
\label{fig4}
\end{figure}

The ground-breaking consequence of this hypothesis is that LSP systems can be used as tracers of extrasolar planetary systems. Fig.~\ref{fig4} shows the on-sky distribution of well-defined sequence~D stars in the Large and Small Magellanic Clouds. If our explanation of the LSP phenomenon is true, the same spatial distribution should have planetary systems in both galaxies. Thus, before the discovery of the first exoplanets outside the Milky Way, we received a tool to analyze their positions in our and other galaxies.

\acknowledgements{This work has been supported by the National Science Centre, Poland, grant MAESTRO No. 2016/22/A/ST9/00009. This publication makes use of data products from the Near-Earth Object Wide-field Infrared Survey Explorer (NEOWISE), which is a joint project of the Jet Propulsion Laboratory/California Institute of Technology and the University of Arizona. NEOWISE is funded by the National Aeronautics and Space Administration.}

\bibliographystyle{ptapap}
\bibliography{paper}

\end{document}